\documentclass{iaus}
\usepackage{graphicx}

\title{The Science of Galaxy Formation}

\author{Gerard Gilmore}

\affiliation{Institute of Astronomy, Madingley Road, Cambridge CB3
  0HA, UK}

\volume{IAUS256}  
\jname{The Galaxy Disk in Cosmological Context}
\editors{J. Andersen, J. Bland-Hawthorn \& B. Nordström eds}

\begin{document}

\maketitle

\begin{abstract}
  Our knowledge of the Universe remains discovery-led: in the absence
  of adequate physics-based theory, interpretation of new results
  requires a scientific methodology. Commonly, scientific progress in
  astrophysics is motivated by the empirical success of the
  ``Copernican Principle'', that the simplest and most objective
  analysis of observation leads to progress. A complementary approach
  tests the prediction of models against observation. In practise,
  astrophysics has few real theories, and has little control over what
  we can observe. Compromise is unavoidable. Advances in understanding
  complex non-linear situations, such as galaxy formation, require
  that models attempt to isolate key physical properties, rather than
  trying to reproduce complexity. A specific example is discussed,
  where substantial progress in fundamental physics could be made with
  an ambitious approach to modelling: simulating the spectrum of
  perturbations on small scales.

\end{abstract}


\section{The Scientific Method}

Astrophysics 
challenges the limits of our scientific methodologies. We have no
control over what Nature allows us to `observe', and much of what we can
observe involves complex non-linear physics. At the
same time, astrophysics challenges the limits of our concepts of ``reality'', so
that our adopted methodolgy is important. Significant astrophysical
queries include the form(s) of the dominant types of matter in the
Universe, the nature of zero-point energy, and, what may be related,
the interpretation of the observed acceleration of the expansion of
the Universe, among other Big Questions. The appropriate scientific
methodology with which to address such questions is itself
problematic: how does one apply what many consider the ``traditional
scientific method'', involving objective analysis of independent
repeated experiments as a test of theory, when the Universe does not
allow us to experiment, in the traditional laboratory physics sense;
when we have no useful predictive theory for much of astrophysics; and
when the nature of the Universe may restrict our observation to only a
very small part of an unobservable larger whole? More specifically, is
the observational test of prediction how science actually
operates?  Is that how astrophysics operates?

The scientific method as popularly conceived is essentially the
application of reason to experience, independent of authority. This
concept has a long and complex evolutionary history, with many notable
figures in its history, from classical Greece, through Ibn Tufayl (see
eg \cite{Cerda-Olmedo}), William of Occam's ``Entia non sunt
multiplicanda praeter necessitatem'', Francis Bacon's discourse in his
``Novum Organum'', Copernicus, Galileo and many more great scientists
and philosophers.  In his paper to the Royal Society in November 1801,
``On the theory of light and colours'', Thomas Young updates Newton's
``Hypotheses non fingo'' in his introduction by ``Although the
invention of plausible hypotheses, independent of any connection with
experimental observations, can be of very little use in promotion of
natural knowledge... '', before introducing what we now know as one of
the great successes and great challenges of the scientific method,
that light behaves as both a wave and a particle.  Niels Bohr, when
becoming a Knight of the Elephant in 1947, adopted the motto
``Contraria sunt Complementa'' (opposites are complementary),
recognising the more general importance of wave-particle duality in
quantum mechanical descriptions of Nature.

This raises two of the more unexpected consequences of application of
the scientific method - is there such a concept as a single ``answer'',
and do the resulting theories describe how the
world ``really is''? How can they, if apparently inconsistent descriptions
are both valid?  Is there such a thing as ``truth'' in science or
Nature? Again to quote Bohr ``It is wrong to think that the task of
physics is to find out how Nature is. Physics concerns what we say
about Nature''. Or, among many hundreds of similar discussions of the
meaning of probability and the role of the observer in quantum
mechanics, von Neumann notes the prime requirement of a model is that
``it is expected to work''. It may well be that abandoning the
classical notion of ``realism'' is the latest step we must take in our
Copernican path to remove observer-specific influence and authority from our
application of reason to some generalised concept of experience ({\textit{cf}}
the discussion in \cite{Leggatt2008}).

Astrophysicists are traditionally proud of their special role in what
is often called the ``Copernican Principle'', the scientific
methodology which applies scepticism to any model of a phenomenon in
which there is a special role for the observer and/or
interpreter. This methodology in astrophysics, and the name, is
derived from empirical ``success''. Removing the special place for
Mankind as the focus of all creation led to a sequence of
models, ranging from Newtonian gravity, through general relativity, to
modern precision cosmology. Along the way the Earth lost its central
place in the Universe, followed by the Sun, then the Milky Way Galaxy.
The concept of absolute time vanished, baryonic matter was dethroned
by dark matter, mass-energy became secondary compared to dark
energy. This last step is a significant extension of the Copernican
Principle. If current speculations on long-term futures in a Universe
dominated by dark energy, Multiverses, and so on, are relevant to
``reality'', the Universe may well be a concept in which what we see,
and what we are, is a temporary fluctuation on what, for most of
space-time, may be very, very different. Cosmic variance becomes not a
consideration but the dominant factor limiting understanding. We, as
observers, may be seeing - or may only be able to see - an extremely
unusual, temporary, microstate, and have no direct knowledge of a
much, much, larger macroscopic ``reality''.

For practising scientists, it is a matter of scientific habit that a
``theory'' which predicts a previously-unobserved phenomenon is
considered supported by experiment. This overstates the case. While
a positive outcome is certainly not neutral, in that the opposite
outcome would lead to quite different reactions, no set
of experiments can ever establish the ``truth'' of any theory. Even if
theory {\textbf{T}} predicts outcome {\textbf{O}}, and {\textbf{O}} is
observed, {\textbf{T}} {\textit{ is not}} proven. If {\textbf{O}} were
outlandish, but observed, it is commonly assumed that {\textbf{T}} is 
more likely to be correct. While a successful test justifies continued
use, and future testing, of that theory, {\textbf{T}} remains
unproven. Supporting the correctness of {\textbf{T}} given the
observation of {\textbf{O}} is the fallacy of ``affirmation of the
consequent'' (cf. \cite{Leggatt2008} for further discussion).

There is no fundamental theory supporting the validity of application
of the ``Copernican Principle''. It is an assumption, whose future
validity, and whose valid range of applications, is unknowable. It may
well be limited. There is certainly no objective justification for its
application in fields beyond those few where it has proven utility.
As an illustration, public reaction to evolutionary biology, and the
scientific realisation that modern, Cro-Magnon man has been painting
caves and doing science for some $10^{-5}$ of the age of the Earth,
remains of considerable complexity, and illustrates well the
difficulty many people have in acting as dispassionate ``Copernican''
observers. There are indeed fields of intellectual enquiry where
objective analysis, independent of the concept of authority, is
inappropriate. A particularly interesting example is the debate in
legal and political circles of the role of the US Supreme Court in
interpreting the US Constitution. Many distinguished legal theorists
insist that a positivist interpretation of what is written, free from
the preferences of specific judges, is most appropriate. Others
disagree. This debate intriguingly combines the concepts of an
authoritative document, and an objective observer and interpreter.
The creativity, sophistication, and continuation, of this debate
illustrates the complexity of the issues. In micro-physics the meaning
and role of the ``observer'' in Young's ``experimental observations''
and the concept of uniqueness, and/or completeness, of possible
observations, have become more complex with developments in quantum
mechanics. The continuing public interest in debating the validity of
string theory as a science (eg \cite{CF07}) is yet another
illustration of both the importance of the questions, and the
incompleteness, or at least complexity, of current interpretations of
the terms ``science'', ``scientific method'', ``theory'' and
``truth''.

\section{The Scientific Method in Galaxy Formation}

With that context, it is perhaps unsurprising that astrophysics
is implemented in a practical approximation to the philosophic
ideal. Many great names in the development of twentieth century
science declared, in essence, ``don't worry too much about the
philosophy, just find, and use, equations which calculate
observables''. Preferably previously un-predicted observables.
In that context, what do we do, and what should we do, in astrophysics.

In practise, we adopt a paradigm, or set thereof, develop it/them in
so far as is possible, testing against, and - hopefully - predicting,
new observables. In that context significant advances have been
made. In astro-particle physics, the interplay between solar structure
models and neutrino astronomy is an exceptional example, as is the
limitation of the numbers and masses of neutrinos from large scale
structure studies.  Steady State cosmology is another exceptional
example -- predictions were made, tested against observations, and the
model found to be inappropriate as a description of the Universe.
Science at its best. Such examples are however rare. Much of
astrophysics either has no {\textit{ab initio}} theory, or involves
complex non-linear physics, so that robust and unique prediction is
impossible. Given our experimental inability to isolate and test
models of individual physical processes, since we are unable to
experiment, we cannot ``test'' the outcome of a theory in astrophysics.

Much of what we do in astrophysics is similar to weather forecasting:
weather forecasts use observations as boundary conditions, implement
the most sophisticated available physics essentially as an
interpolation (in space, in time,...), exploit heroic achievements in
computing, and extrapolate the observables to other places and
times. Sometimes this is accurate, sometimes not, in which case the
differences between prediction and observation are analysed to allow
the forecasting system to be improved.  Eventually, given enough data,
and enough complexity in the model, weather forecasts will become,
asymptotically, as accurate as the predictability of the system
allows. They will reach a physical accuracy limit. But no weather
forecast can ever be ``right'' or ``wrong'', in the sense that a
scientific theory can be. A forecast may be accurate, or less
accurate. It is unlikely even that any forecasting system could ever
be unique, since there may well be many physical processes whose
effects are comparable in amplitude to measurement error. A system
with considerable complexity, and inevitable approximation, will
invariably have many statistically-indistinguishable solution maxima.

Coming specifically to models of galaxy formation, we have a similar
situation.  There is an interesting distinction between (some) Galaxy
(ie, Milky Way) models and (some) galaxy (ie generic)
models. Substantial progress is being made in development of specific
models of the Milky Way Galaxy, particularly in preparation for Gaia.
Gaia will produce information from which we expect to determine the
current state of the Milky Way Galaxy in some detail, and hence to
deduce something of how the Milky Way in particular, and,
{\textit{modulo}} cosmic variance, disk galaxies in general, formed
and evolved. A systematic approach to modelling, analysing and
interpreting the anticipated Gaia data is underway. The adopted
strategy is to proceed through a sequence of models of increasing
complexity, guided at each stage by analysis of mis-matches betwen the
current model and available simulations, real data and on-going
surveys, such as RAVE (see eg \cite{B02}, as one example of the many
underway). This process is intended to develop what is essentially a
tool-kit for investigation of the Gaia dataset, and hence the Milky
Way Galaxy. This modelling approach is, in a real sense, equivalent to a
laboratory experiment, rather than being development of a theory.

Formation models for galaxies in general are very different in
approach and ambition. They adopt analyses of the properties of the
early universe, derived from observations of the cosmic microwave
background, and supplementary data, as boundary conditions. These
boundary conditions are unconstrained by observations on small scales,
and so are extrapolated [usually as a simple power-law spectrum of
fluctuations] down to as-yet unobserved physical length scales. This
extrapolated set of boundary conditions is then evolved forward in
time, requiring considerable sophistication and heroic achievements in
computing.  Approximations to the behaviour of baryons, and hence the
properties of most observables, are then added.  Comparison with
observations of real galaxies, when made, has so far invariably
identified gross discrepancies, indicating perhaps that more complex
baryon physics is needed. Or different physics: perhaps the
extrapolation of the observational boundary conditions is
inappropriate? Unfortunately, analysis and interpretation of the
predictions of these inevitably highly idealised models is complex.
 
In order to calculate ``observables'' {\textit{ad hoc}} prescriptions
for the key baryonic physics must be added by hand. Star formation,
chemical elements, black holes and so on are added using some
observationally motivated recipe. After unsuccessful comparison to
observation, the complexity is increased, including both plausibly
anticipated and some quite {\textit{ad hoc}} effects -- bias,
scale-dependant bias, feedback, AGN feedback, ...  etc, are
included. The complexities of ``post-formation'' dynamical evolution
(or even survival) must all be approximated. And so on. Considerable
current effort is involved in adjusting the non-linear aspects of the
baryonic physics to try to regain consistency with observation.
Consistency with observation is not a natural feature of extant models
of galaxy formation.

The development of the currently available sophisticated galaxy models
is a powerful and extremely impressive achievement. Is it developing a
theory?  There is no {\textit{ab initio}} theory, no first-principles
calculation, of many of the physical processes.  It is feasible that a
model can be identified, with eventual sufficient complexity, which is
able to reproduce all extant observables.  This will not be a
theory. It will never be ``right'' or ``wrong''.  Until key parameter
space is investigated, no model will even be unique within its limited
starting points and methodology. That is, there is a fundamental
distinction between development of a model/tool-kit which is
appropriate to investigate Gaia-like data sets, and modelling galaxy
formation from linear perturbations early in the Universe. The latter
models can never be compared to data, except after `processing'
through complex non-linear processes, which are themselves neither
understood nor quantified.

So is there any point in devoting effort to building complex models of
galaxy formation, when they are inherently untestable and not unique?
Yes! In fact, such modelling can, or could, already be used for
important investigations of some key assumptions in general
astrophysics and cosmology. Galaxy formation models, given their
present (impressive) sophistication, are valuable tools to investigate
hypotheses. As yet, however, the models are incapable of testing
hypotheses as complex as the formation of a galaxy.  Appropriate
hypotheses to test are more fundamental than the highly specific
challenge of adding complexity to a recipe to become not-inconsistent
with extant observations. Galaxy formation models are, as yet, not
very helpful tools to determine the details of the complex mix of
non-linear physics which describes the evolution of baryons and dark
matter on small scales in a galaxy. Galaxy formation models could
however, if applied appropriately, be a very valuable tool-kit to
investigate much more fundamental physics.

\begin{figure}[h!]
\begin{center}
 \includegraphics[width=4.4in]{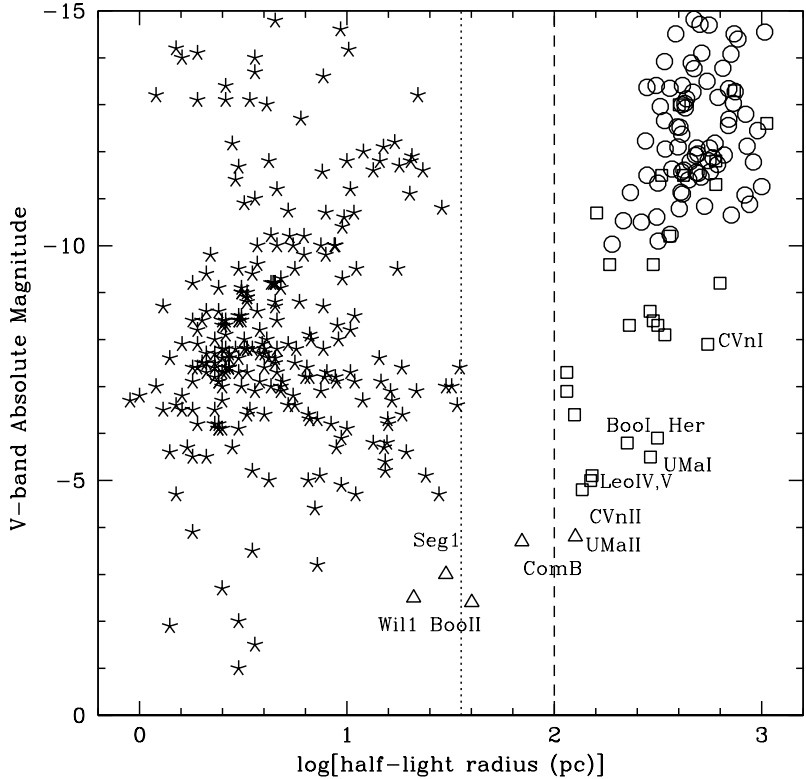} 
 \caption{The relation between absolute luminosity and luminous
   half-light radius for small stellar systems in the local Universe.
   Globular clusters from several host galaxies, Ultra Compact Dwarfs,
    and galactic nuclei star clusters, are represented as
   asterisks. Local Group dSph galaxies, with the most newly
   discovered identified by name, are shown as open squares.  Galaxies
   from the Local Volume survey of \cite{S08} are shown as open
   circles. Milky Way satellites of unknown equilibrium status are
   shown as open triangles (see Fig~2). All equilibrium galaxies have
   half-light radii larger than the minimum size line at 100pc. All
   apparently purely stellar systems have half-light radii smaller
   than about 30pc. Further details are in \cite{G07}. }
   \label{fig1}
\end{center}
\end{figure}

\begin{figure}[h!]
\begin{center}
 \includegraphics[width=3.4in,angle=-90]{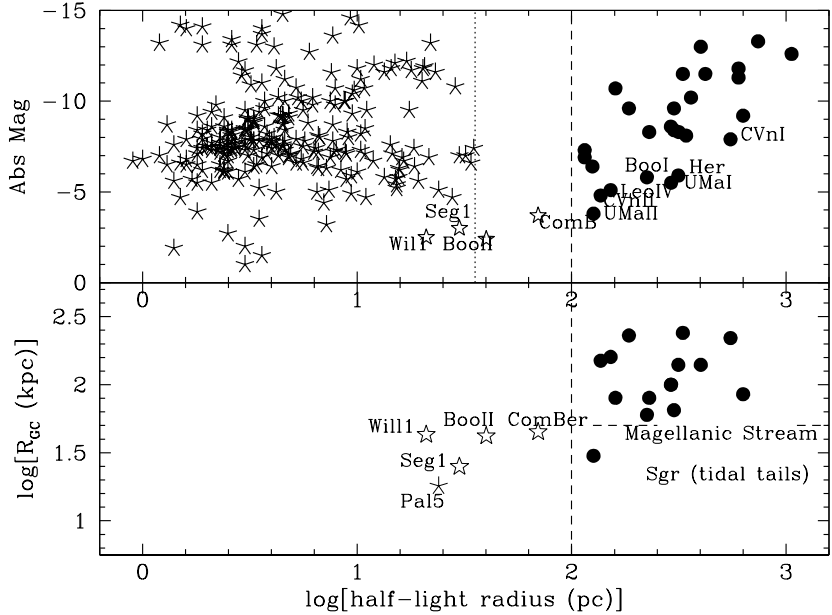} 
 \caption{The observed structural properties of the smallest galaxies
   and stellar systems. Top: relation between absolute luminosity and
   luminous half-light radius. Globular clusters, Ultra Compact
   Dwarfs, and galactic nuclei star clusters, are represented
   as asterisks. dSph galaxies, with the most newly discovered
   identified, are shown as solid points. Objects of unknown
   equilibrium status are shown as open stars. All objects which show
   robust evidence for dark matter halos have half-light radii larger
   than the minimum size line at 100pc. All stellar systems have
   half-light radii smaller than about 30pc, and none shows evidence
   for dark matter. Bottom: the uncertain dynamical state of the
   intermediate objects is emphasised by considering size as a
   function of Galacto-centric distance. All uncertain objects are in
   a region where Galactic tides are expected to be important, and so
   may have time-dependent structures. Further details are in
   \cite{G07}. }
   \label{fig2}
\end{center}
\end{figure}

\subsection{What galaxy formation models could tell us}

There are fundamental questions in physics, and in $\Lambda$CDM
cosmology, which {\textit{are}} best addressed using galaxy formation
models. Applications to the fundamental properties of neutrinos are
mentioned above. To give just one more important example, the standard
model of particle physics is known to be incomplete. Extension to a
more general theory requires guidance from observations. Recently,
most of these new observations have come from astrophysics - neutrino
masses, baryogenesis, matter-anti-matter asymmetry, the dominance of
dark matter, the importance of dark energy, are among this list. The
minimal super-symmetric extension of the particle physics standard
model, which does not even encompass all the complexity required to
address all the items on this list, has more than 120 free
parameters. Hopefully, in the near future, CERN will advance
measurement of aspects of the parameter space. Astrophysics has done
so -- limits on neutrino masses from large scale structure are a
superb example -- but can do much more, including invetsigating
aspects of physics at a more fundamental scale than is possible with
accelerators.

Perhaps the best and most immediate example is in testing the
small-scale extension of the spectrum of perturbations. At present,
$\Lambda$CDM models adopt the spectrum of perturbations from analysis
of CMB and other observations, and extend this to zero scale. The
extension is unphysical, in being ultraviolet divergent. Suppression
of the divergence is provided essentially by numerical smoothing
(``finite resolution'') in cosmological simulations. It is unlikely
that Nature does it that way. Rather, the small-scale power spectrum
may well be where astroparticle physics comes into action on
observable scales. Testing this is arguably much more interesting than
is applying ingenuity to fine-tune outcomes of the models to make them
not-inconsistent with already known observations.

\begin{figure}[h!]
\begin{center}
 \includegraphics[width=3.4in]{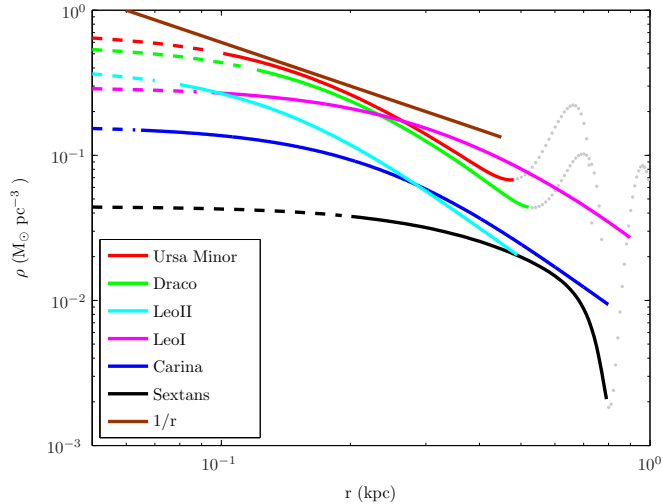} 
 \caption{Derived internal mass profiles for the well-studied dSph
   galaxies. In each case a cored dark matter distribution is
   preferred by the kinematic data, with a scale radius comparable to
   that of the luminous scale shown in Figures 1 \& 2. The similarity of
   this scale in all cases studied implies it is an inherent property
   of dark matter itself. Further details are in \cite{G07}.}
   \label{fig3}
\end{center}
\end{figure}

A huge literature is available considering the implications of
specific possibilities and elementary particle types in
astrophysics. \cite{OS2003} provide an accessible summary of the
interplay between small-scale structure and classes of
physically motivated explanations for suppression of the ultraviolet
divergence. The key is that physical effects may be expected on scales
of the smallest galaxies, the dSph. It is perhaps not coincidence that
it is on these scales that the bottom-up galaxy formation models have
proven to be most inconsistent with observations. The list of relevant
issues is long, and well-known. Examples from a long list include the
existence of old red massive galaxies, the Tully-Fisher relation being
in place by redshift unity, the frequency of large old cold disks in
galaxies, and the satellite problem. Such a long list of observations
all inconsistent with apparently fundamental features of galaxy
formation models suggests two approaches. In one approach, new complex
physics (``feedback'') must be added, to ``improve'' agreement with
observation. The appearances are to be saved.  In another, common
assumptions in the galaxy simulations could be examined further.

The recent observational status of the small-scale problem is
described in \cite{G07}. An updated summary is presented in figures 1,
2, and 3 here.  There are two key results: the lowest luminosity
galaxies are all very dark-matter dominated, and all have an
equilibrium minimum half-light optical radius of $>100$pc. The
largest star clusters have half-light radius less than 30pc. No
equilibrium objects are known in the local Universe with half-light
size between 30pc and 100pc (Figures~1,2).  Dynamical studies of
stellar kinematics in very low-luminosity galaxies all prefer a
dark-matter distribution which is cored, with a mass scale length
comparable to the luminosity scale length (Figure~3).  Standard
assumptions adopted in simulations of galaxy formation have no
presumed physical scale (ie, the UV divergence), so featureless smooth
distributions are a natural feature. A specific length scale is not
anticipated, but is seen. This physical scale seems to be special.

It may well be that we are discovering a physics-based solution to the
medley of challenges to galaxy formation models: the divergence of the
small-scale extrapolation of the perturbation spectrum is at
fault. The physics of the mix of dark matter particles may be the
explanation. While it will require considerable ingenuity to extend the
resolution of numerical simulations to handle such small scales
reliably, this is an example where simulations could explore the
effect of the power spectrum, and so investigate a new regime. That is
a physics experiment which really can test a physical theory: using
observations of galaxies as a guide, is there an astrophysically
observable physical scale at which the power spectrum converges? What
is the sensitivity of predictions of galaxy formation to the assumed
boundary conditions on small scales? What classes of elementary
particles must then make up much of the dark matter on small scales?

\subsection{A constraint on early substructure}

Among the most direct measures of the size and location of early star
formation is scatter in chemical element ratios (see papers here by
Wyse, by Nissen, and others). Small scatter requires  a large and
well-mixed star-forming region, which has an independent existence for
sufficiently long to self-enrich. Careful quantification of the
scatter in element ratios as a function of [M/H] clearly can count the
number of star-forming events in the early Galaxy directly. This is
of course well known, and has been so for many years. An interesting
extension of this analysis can be applied to the light elements
Beryllium (and Lithium) which are made, fully for Be, partly for Li,
by cosmic ray spallation, probably with CNO nuclei as cosmic ray
primaries spallating onto H-nuclei (cf Pasquini, this meeting, and
\cite{GGEN92}). This spallation involves very high-energy heavy
nuclei, probably accelerated by the same supernovae in which they were
created. Such high-energy particles have a very long mean free
path. They cannot be retained inside a small or short-lived star
forming event, such as a small, transient, dark matter halo. Thus,
inevitably, any stars formed in such small halos will have little or
no Be. Their Li abundances will also provide a robust determination of
the relative contributions of BBN and later spallation to their Li
abundances. Determination of the range of Be abundances in field halo
stars, and -- ideally -- in either a low-mass dSph galaxy or a
verifiable kinematic stream, will provide extremely interesting
constraints on the range of places where early star formation occurred.

\section{An historical lesson}

This meeting celebrates the centenary of Bengt Str\"omgren. Openness
to the implications of observations, and an ability to move beyond
preconceptions, was one of his great attributes. My personal example
involved his long-standing research interest in the formation and
evolution of the Galactic disk. After decades of work, developing the
Str\"omgren photometric system, and acquiring vast data sets, Bengt
Str\"omgren, in retirement(!!) was near to finalising his major study
of the distribution of stellar ages and abundances near the Sun. In
1983 Gilmore \& Reid announced their discovery of the Galactic thick
disk. The thick disk stellar population is old and relatively
metal-rich [\cite{GW85}], with a main-sequence turn-off to the red of
the F-star range which was at the time being studied in Str\"omgren's
survey.  Seeing this result, Bengt Str\"omgren invited me to visit,
rapidly pursuaded himself that his extant survey was biassed by being
based on a too-restrictive assumption on the past age-metallicity
relation, and so extended his survey to include redder stars. He did
this knowing that he might well not live to see the outcome of his
lifetime research project. An impressive example of scientific
objectivity, indeed. Fortunately, his colleagues worked hard, the
weather was good, so Bengt Str\"omgren was able to present the first
results of his expanded survey in his last scientific paper,
\cite{Strom87}.

\end{document}